\documentclass[aps,print,showpacs,notitlepage,12pt,onecolumn]{revtex4}%
\usepackage{amsfonts}
\usepackage{amsmath}
\usepackage{amssymb}
\usepackage{graphicx}%
\begin{document}
\title{Exact soliton solution of Spin Chain with a external magnetic field in linear
wave background }
\author{Qiu-Yan Li, Zheng-Wei Xie }
\affiliation{Department of Physics, Sichuan Normal University, Chengdu 610068, China}
\author{Lu Li, Z. D. Li, J. Q. Liang}
\affiliation{Institute of Theoretical Physics and Department of Physics, Shanxi University,
Taiyuan 030006, China}

\begin{abstract}
Employing a simple, straightforward Darboux transformation we construct exact
N-soliton solution for anisotropic spin chain driven by a external magnetic
field in linear wave background. As a special case the explicit one- and
two-soliton solution dressed by the linear wave corresponding to magnon in
quantum theory is obtained analytically and its property is discussed in
detail. The dispersion law, effective soliton mass, and the energy of each
soliton are investigated as well. Our result show that the stability criterion
of soliton is related with anisotropic parameter and the amplitude of the
linear wave.

\end{abstract}

\maketitle

\section{Introduction}

The concept of soliton in spin chain which exhibits both coherent and chaotic
structures depending on the nature of the magnetic
interactions\cite{Trullinger,Kivshar,Ablowitz,Fokas} has received considerable
attention for decades. Soliton in quasi one-dimensional magnetic systems is no
longer a theoretical concept but can be probed by neutron inelastic scattering
\cite{Kjems78,Boucher90}, Mossbauer linewidth measurements \cite{Thiel81}, and
electron spin resonance \cite{Asano00}. One of the integrable models for spin
chain is Landau-Lifshitz equation \cite{landau} which has been studied by a
variety of techniques such as inverse scattering
transformation\cite{Ablowitz,Takhtajan,Chen}, Darboux transformation
\cite{Huang}, Riemann-Hilbert approach\cite{liu1}, etc. Exact soliton
solutions are reported for the isotropic
\cite{Takhtajan,Tjon,Fogedby,Laksmanan,Shimizu} and anisotropic
\cite{Chen,Huang,Mikeska,Li,liu1} spin chains as well. The other integrable
model is a type of Nonlinear-Schr\H{o}dinger equation studied in Ref.
\cite{Ablowitz}. It should be noted that all these solutions are obtained in
the ground state background. The exact soliton solutions in linear wave
background corresponding to magnon in quantum theory\cite{Majlis}\ has
received no attention yet. The main goal of this paper is to search for new
solutions of spin chain driven by external magnetic field in linear wave
background. We obtain exact solution of N-soliton train in terms of a simple,
straightforward Darboux transformation\cite{Matveev,C.H. Gu}. Particularly the
time-evolution of one and two-soliton is analyzed in terms of the general solution.

The outline of this paper is organized as follows: In Sec. II the Darboux
transformation is explained in detail and the general N-soliton solution is
obtained. In Sec. III we discuss a special case. Exact one-soliton solution in
linear wave background is obtained. The dispersion law and soliton energy are
investigated as well. Sec. IV is devoted to general two-soliton solution and
soliton collisions. Finally, our concluding remarks are given in Sec. V.

\section{Exact solution of N-soliton train}

Our starting Hamiltonian describing the anisotropic spin chain with a external
magnetic field can be written as
\begin{align}
\hat{H}  &  =-g\mu_{B}B\sum_{j}\hat{S}_{j}^{z}-J^{\prime}\sum_{j,\varepsilon
}\hat{S}_{j}^{z}\hat{S}_{j+\varepsilon}^{z}\nonumber\\
&  -\frac{1}{2}J\sum_{j,\varepsilon}\left(  \hat{S}_{j}^{+}\hat{S}%
_{j+\varepsilon}^{-}+\hat{S}_{j}^{-}\hat{S}_{j+\varepsilon}^{+}\right)  ,
\label{ham1}%
\end{align}
where $\widehat{S}_{j}\equiv(\widehat{S}_{j}^{x},\widehat{S}_{j}^{y}%
,\widehat{S}_{j}^{z})$ with $j=1,2,...,N$ are spin operators, $J^{\prime}>J$
$>0$ is the pair interaction parameter, $g$ the Lande factor and $\mu_{B}$ is
the Bohr magneton, $B$ is the external magnetic field. With the help of
Holstein-Primokoff \cite{Holstein} transformation $\hat{S}_{j}^{z}%
=S-a_{j}^{\dagger}a_{j}$, $\hat{S}_{j}^{+}\approx\sqrt{2S}\left(  1-\frac
{1}{4S}a_{j}^{\dagger}a_{j}\right)  a_{j}$, $\hat{S}_{j}^{-}\approx\sqrt
{2S}a_{j}^{\dagger}\left(  1-\frac{1}{4S}a_{j}^{\dagger}a_{j}\right)  $, the
Hamiltonian in Eq. (\ref{ham1}) reduces to
\begin{align*}
\hat{H}  &  =-2J^{\prime}S^{2}N-g\mu_{B}BSN+g\mu_{B}B\sum_{j}a_{j}^{\dagger
}a_{j}\\
&  +S\sum_{j,\varepsilon}\left\{  J^{\prime}(a_{j}^{\dagger}a_{j}%
+a_{j+\varepsilon}^{\dagger}a_{j+\varepsilon})-J(a_{j}a_{j+\varepsilon
}^{\dagger}+a_{j}^{\dagger}a_{j+\varepsilon})\right\} \\
&  -J^{\prime}\sum_{j,\varepsilon}a_{j}^{\dagger}a_{j}a_{j+\varepsilon
}^{\dagger}a_{j+\varepsilon}+\frac{1}{4}J\sum_{j,\varepsilon}\left[
{}\right.  a_{j}^{\dagger}a_{j+\varepsilon}^{\dagger}a_{j+\varepsilon
}a_{j+\varepsilon}\\
&  +a_{j+\varepsilon}^{\dagger}a_{j}^{\dagger}a_{j}a_{j}+a_{j}^{\dagger}%
a_{j}^{\dagger}a_{j}a_{j+\varepsilon}+a_{j+\varepsilon}^{\dagger
}a_{j+\varepsilon}^{\dagger}a_{j+\varepsilon}a_{j}\left.  {}\right]
\end{align*}
where the boson operators $a_{j}$ are assumed to satisfy the usual commutation
relation $[a_{j},a_{j^{\prime}}^{\dagger}]=\delta_{jj^{\prime}}$,
$[a_{j}^{\dagger},a_{j^{\prime}}^{\dagger}]=[a_{j},a_{j^{\prime}}]=0$. The
equation of motion for the operator $a_{j}$ on the nth site is $i\hbar
\frac{\partial}{\partial t}a_{j}=[a_{j},\hat{H}]$. At low temperatures, the
operator $a_{j}$ can be treated as a classical vector such that $a_{j}%
\rightarrow a(x.t)$. So that the equation of motion in a continuum spin chain
under a magnetic field can be obtained as a Nonlinear-Schr\H{o}dinger type:%

\begin{equation}
-i\frac{\partial}{\partial t}a{}=\beta_{0}\frac{\partial^{2}}{\partial x^{2}%
}a{}+2\beta_{0}\beta_{1}^{2}a\left\vert a\right\vert ^{2}+2\beta_{2}a,
\label{spin1}%
\end{equation}
where
\[
\beta_{0}=\frac{2JS}{\hbar},\beta_{1}=\sqrt{\frac{J^{^{\prime}}-J}{JS}}%
,\beta_{2}=-\frac{g\mu_{B}B+4\left(  J^{^{\prime}}-J\right)  S}{2\hbar},
\]
here $J^{^{\prime}}>J>0$(easy-axis). In this paper we will present a
systematic method to construct general expressions of one- and two-soliton
solutions embedded in a linear wave background for Eq. (\ref{spin1}) and their
novel properties.

By employing Ablowitz-Kaup-Newell-Segur technique one can construct the linear
eigenvalue problem for Eq. (\ref{spin1}) as follows
\begin{equation}
\psi_{x}=U\psi,\text{ }\psi_{t}=F\psi, \label{Laxpair}%
\end{equation}
where $\psi=\left(
\begin{array}
[c]{l}%
\psi_{1}\\
\psi_{2}%
\end{array}
\right)  $, $U$ and $F$ can be given in the forms
\begin{align*}
U  &  =\lambda\sigma_{3}+P,\\
F  &  =i\left(  2\lambda^{2}\beta_{0}+\beta_{2}\right)  \sigma_{3}%
+i2\lambda\beta_{0}P-i\beta_{0}\left[  P^{2}+P_{x}\right]  \sigma_{3},
\end{align*}
with%
\[
\sigma_{3}=\left(
\begin{array}
[c]{ll}%
1 & \text{ \hspace{0in}}0\\
0 & -1
\end{array}
\right)  ,P\left(  x,t\right)  =\left(
\begin{array}
[c]{ll}%
0 & \beta_{1}a\\
-\beta_{1}\overline{a} & 0
\end{array}
\right)  ,
\]
and the overbar denotes the complex conjugate. Thus Eq. (\ref{spin1}) can be
recovered from the compatibility condition $U_{t}-F_{x}+\left[  U,F\right]
=0$. Based on the Lax pair (\ref{Laxpair}), we can obtain general one- and
two-soliton solution embedded in a linear wave background by using a
straightforward Darboux transformation\cite{Matveev,C.H. Gu}.

Consider the following transformation
\begin{equation}
\Psi=\left(  \lambda I-K\right)  \psi,\text{ }K=H\Lambda H^{-1},\text{
}\Lambda=\text{diag}\left(  \lambda_{1},\lambda_{2}\right)  , \label{Darboux1}%
\end{equation}
where $H$ \hspace{0in}is a nonsingular matrix which satisfies
\begin{equation}
H_{x}=\sigma_{3}H\Lambda+PH. \label{Darboux2}%
\end{equation}
Letting
\begin{equation}
\Psi_{x}=U_{1}\Psi, \label{Darboux3}%
\end{equation}
where $U_{1}=\lambda\sigma_{3}+P_{1}$, $P_{1}=\left(
\begin{array}
[c]{ll}%
0 & \beta_{1}a_{1}\\
-\beta_{1}\overline{a}_{1} & \text{ \hspace{0in} \hspace{0in} }0
\end{array}
\right)  $, and with the help of Eqs. (\ref{Laxpair}), (\ref{Darboux1}) and
(\ref{Darboux2}), we obtain the Darboux transformation for Eq. (\ref{spin1})
from Eq. (\ref{Darboux3}) in the form
\begin{equation}
P_{1}=P+\left[  \sigma_{3},K\right]  \text{.} \label{Darboux4}%
\end{equation}
It is easy to verify that, if $\psi=\left(
\begin{array}
[c]{l}%
\psi_{1}\\
\psi_{2}%
\end{array}
\right)  $ is a eigenfunction of Eq. (\ref{Laxpair}) with eigenvalue
$\lambda=\lambda_{1}$, then $\left(
\begin{array}
[c]{l}%
-\overline{\psi}_{2}\\
\overline{\psi}_{1}%
\end{array}
\right)  $ is also the eigenfunction, however with eigenvalue $-\overline
{\lambda}_{1}$. Thus if taking
\begin{equation}
H=\left(
\begin{array}
[c]{ll}%
\psi_{1} & -\overline{\psi}_{2}\\
\psi_{2} & \text{ }\overline{\psi}_{1}%
\end{array}
\right)  ,\Lambda=\left(
\begin{array}
[c]{ll}%
\lambda_{1} & \text{ }0\\
0 & -\overline{\lambda}_{1}%
\end{array}
\right)  , \label{Darboux5}%
\end{equation}
which ensures that Eq. (\ref{Darboux2}) is held, we can obtain
\begin{equation}
K_{sl}=-\overline{\lambda}_{1}\delta_{sl}+\left(  \lambda_{1}+\overline
{\lambda}_{1}\right)  \frac{\psi_{s}\overline{\psi}_{l}}{\psi^{T}%
\overline{\psi}},\text{ }s,l=1,2, \label{Darboud6}%
\end{equation}
where $\psi^{T}\overline{\psi}=\left\vert \psi_{1}\right\vert ^{2}+\left\vert
\psi_{2}\right\vert ^{2},$ and Eq. (\ref{Darboux4}) becomes
\begin{equation}
a_{1}=a+\frac{2}{\beta_{1}}\left(  \lambda_{1}+\overline{\lambda}_{1}\right)
\frac{\psi_{1}\overline{\psi}_{2}}{\psi^{T}\overline{\psi}}, \label{Darboux}%
\end{equation}
where $\psi=\left(  \psi_{1},\psi_{2}\right)  ^{T}$ is the eigenfunction of
Eq. (\ref{Laxpair}) corresponding to the eigenvalue $\lambda_{1}$ for the
solution $a$. Thus by solving the Eq. (\ref{Laxpair}) which is a first-order
linear differential equation, we can generate a new solution $a_{1}$ of the
Eq. (\ref{spin1}) from a known solution $a$ which is usually called
\textquotedblleft seed\textquotedblright\ solution.

To obtain exact $N$-order solution of Eq. (\ref{spin1}), we firstly rewrite
the Darboux transformation in Eq. (\ref{Darboux}) as in the form
\begin{equation}
a_{1}=a+\frac{2}{\beta_{1}}\left(  \lambda_{1}+\overline{\lambda}_{1}\right)
\frac{\psi_{1}\left[  1,\lambda_{1}\right]  \overline{\psi}_{2}\left[
1,\lambda_{1}\right]  }{\psi\left[  1,\lambda_{1}\right]  ^{T}\overline{\psi
}\left[  1,\lambda_{1}\right]  }, \label{Oneso}%
\end{equation}
where $\psi\left[  1,\lambda\right]  =\left(  \psi_{1}\left[  1,\lambda
\right]  ,\psi_{2}\left[  1,\lambda\right]  \right)  ^{T}$ denotes the
eigenfunction of Eq. (\ref{Laxpair}) corresponding to eigenvalue $\lambda$.
Then repeating above the procedure for $N$ times, we can obtain the exact
$N$-order solution
\begin{equation}
a_{N}=a+\frac{2}{\beta_{1}}\sum_{m=1}^{N}(\lambda_{m}+\overline{\lambda}%
_{m})\frac{\psi_{1}[m,\lambda_{m}]\overline{\psi}_{2}[m,\lambda_{m}]}%
{\psi\lbrack m,\lambda_{m}]^{T}\overline{\psi}[m,\lambda_{m}]},
\label{Multiso}%
\end{equation}
where
\begin{align*}
\psi\left[  m,\lambda\right]   &  =\left(  \lambda-K\left[  m-1\right]
\right)  \cdots\left(  \lambda-K\left[  1\right]  \right)  \psi\left[
1,\lambda\right]  ,\\
K_{sl}\left[  j^{\prime}\right]   &  =-\overline{\lambda}_{j^{\prime}}%
\delta_{sl}+\left(  \lambda_{j^{\prime}}+\overline{\lambda}_{j^{\prime}%
}\right)  \frac{\psi_{s}\left[  j^{\prime},\lambda_{j^{\prime}}\right]
\overline{\psi}_{l}\left[  j^{\prime},\lambda_{j^{\prime}}\right]  }%
{\psi\left[  j^{\prime},\lambda_{j^{\prime}}\right]  ^{T}\overline{\psi
}\left[  j^{\prime},\lambda_{j^{\prime}}\right]  },
\end{align*}
here $\psi\left[  j^{\prime},\lambda\right]  $ is the eigenfunction
corresponding to $\lambda_{j^{\prime}}$ for $a_{j^{\prime}-1}$ with
$a_{0}\equiv a$ and $s,l=1,2$, $j^{\prime}=1,2,\cdots,m-1$, $m=2,3,\cdots,N$.
Thus if choosing a \textquotedblleft seed\textquotedblright\ as the basic
initial solution, by solving linear characteristic equation system
(\ref{Laxpair}), one can construct a set of new solutions for Eq.
(\ref{spin1}) by employing the formula (\ref{Multiso}).

As an example, we give the exact expression of one- and two-soliton solutions
in a linear wave background for Eq. (\ref{spin1}) respectively and analyze its
properties. For this propose, we take the initial \textquotedblleft
seed\textquotedblright\ as $a=A_{c}e^{ik_{c}x-i\omega_{c}t}$ which is a
linear-wave solution and satisfies the nonlinear dispersion relation
$\omega_{c}=\beta_{0}\left(  k_{c}^{2}-2\beta_{1}^{2}A_{c}^{2}\right)
-2\beta_{2}$, where $A_{c}$ and $\omega_{c}$ are the arbitrary real constants.
Substituting the initial seed into (\ref{Laxpair}) and solving the linear
equation, by tedious calculation we obtain the expression of eigenfunction
corresponding to eigenvalue $\lambda$ in the form
\begin{align}
\psi_{1}  &  =-\beta_{1}A_{c}C_{1}\exp\Theta_{1}+L_{1}C_{2}\exp\Theta
_{2},\nonumber\\
\psi_{2}  &  =L_{1}C_{1}\exp\left(  -\Theta_{2}\right)  -\beta_{1}A_{c}%
C_{2}\exp\left(  -\Theta_{1}\right)  , \label{Laxsolution}%
\end{align}
where
\begin{align*}
\Theta_{1}  &  =\frac{1}{2}i\varphi+iM\left(  x-\gamma t\right)  ,\text{
}\Theta_{2}=\frac{1}{2}i\varphi-iM\left(  x-\gamma t\right)  ,\\
L  &  =\lambda-i\frac{k_{c}}{2}-iM,\text{ }M=\frac{1}{2}\sqrt{\left(
k_{c}+i2\lambda\right)  ^{2}+4\beta_{1}^{2}A_{c}^{2}},\\
\gamma &  =\left(  k_{c}-i2\lambda\right)  \beta_{0},\text{ }\varphi
=k_{c}x-\omega_{c}t.
\end{align*}
here the parameters $C_{1}$ and $C_{2}$ are the arbitrary complex constants.
Following the expression of eigenfunction and with the aid of the formulas
(\ref{Multiso}), we can obtain exactly the one- and two-soliton solutions
embedded in a linear wave background, respectively.

\section{One-soliton solution embedded in linear wave background}

Taking the spectral parameter $\lambda=\lambda_{1}\equiv\frac{A_{s,1}}%
{2}+i\frac{k_{s,1}}{2}$ in Eq. (\ref{Laxsolution}) and substituting them into
Eq. (\ref{Oneso}), we obtain the one-soliton solution in the linear wave
background as follows
\begin{align}
a_{1}  &  =e^{i\varphi}\left\{  {}\right.  A_{c}+\frac{A_{s,1}}{\beta
_{1}\Delta_{1}}\left[  {}\right.  \beta_{1}^{2}A_{c}^{2}e^{i\Phi_{1}%
}+\left\vert L_{1}\right\vert ^{2}e^{-i\Phi_{1}}\nonumber\\
&  -2\beta_{1}A_{c}\left(  \operatorname{Re}L_{1}\cosh\theta_{1}%
+i\operatorname{Im}L_{1}\sinh\theta_{1}\right)  \left.  {}\right]  \left.
{}\right\}  \label{S1}%
\end{align}
where%
\begin{align*}
\Delta_{1}  &  =\left(  \left\vert L_{1}\right\vert ^{2}+\beta_{1}^{2}%
A_{c}^{2}\right)  \cosh\theta_{1}-2\beta_{1}A_{c}\left(  \operatorname{Re}%
L_{1}\right)  \cos\Phi,\\
\theta_{1}  &  =2\operatorname{Im}M_{1}\left(  x-V_{1}t\right)  -x_{0},\text{
}\\
\Phi_{1}  &  =2\operatorname{Re}M_{1}\left(  x-V_{2}t\right)  -\varphi_{0},\\
V_{1}  &  =\frac{\operatorname{Im}\left(  M_{1}\gamma_{1}\right)
}{\operatorname{Im}M_{1}},\text{ }V_{2}=\frac{\operatorname{Re}\left(
M_{1}\gamma_{1}\right)  }{\operatorname{Re}M_{1}},\text{ }\varphi
=k_{c}x-\omega_{c}t,
\end{align*}
with the parameters $M_{1}=\frac{1}{2}\sqrt{(k_{c}+i2\lambda_{1})^{2}%
+4\beta_{1}^{2}A_{c}^{2}}$, $L_{1}=\lambda_{1}-ik_{c}/2-iM_{1}$, and
$\gamma_{1}=\left(  k_{c}-i2\lambda_{1}\right)  \beta_{0}$. The parameters
$x_{0}$ and $\varphi_{0}$ represent the initial center position and initial
phase, which are determined by $x_{0}=-\ln\left\vert C_{2}/C_{1}\right\vert $,
and $\varphi_{0}=\arg\left(  C_{2}/C_{1}\right)  ,$ respectively$,$ where
$C_{1},C_{2}$ are the arbitrary complex constants. It is worth to note that
because $x_{0}$ and $\varphi_{0}$ are determined by the value $C_{2}/C_{1},$
they in fact depend on only one arbitrary complex parameter. Not loss of
generality, we take $C_{1}=1$.

The exact solution $a_{1}$ in Eq. (\ref{S1}) describes a soliton solution of
anisotropic spin chain embedded in a linear-wave background with the soliton
amplitude $\frac{A_{s,1}}{\beta_{1}}$, the width $\frac{1}{2\operatorname{Im}%
M_{1}}$, the wavenumber $k_{1}=2\operatorname{Re}M_{1}$, the frequency
$\Omega_{1}=2\operatorname{Re}\left(  M_{1}\gamma_{1}\right)  $, and the
envelope velocity $V_{1}=\operatorname{Im}\left(  M_{1}\gamma_{1}\right)
/\operatorname{Im}M_{1}$. As the linear-wave amplitude vanishes, namely
$A_{c}=0$, this solution in Eq. (\ref{S1}) reduces to the solution in the from%
\begin{equation}
a_{1-\text{sol}}=\frac{A_{s}e^{i\left[  k_{s,1}x-\Omega_{s,1}t+\varphi
_{0}\right]  }}{\beta_{1}\cosh\left[  A_{s,1}\left(  x-V_{s,1}t-x_{0}^{\prime
}\right)  \right]  }, \label{Br}%
\end{equation}
where $x_{0}^{\prime}$ is determined by $x_{0}^{\prime}=\frac{-1}{A_{s,1}}%
\ln\left\vert C_{2}\right\vert $. The solution $a_{1-\text{sol}}$ in Eq.
(\ref{Br}) describes a bright soliton solution with maximal amplitude
$\frac{A_{s,1}}{\beta_{1}}$, the width $\frac{1}{A_{s,1}}$, envelope velocity
$V_{s,1}=2\beta_{0}k_{s,1}$, and center position $x_{0}^{\prime}$. The
frequency $\Omega_{s,1}=\frac{1}{\hbar}[g\mu_{B}B+4\left(  J^{^{\prime}%
}-J\right)  S-2JSA_{s,1}^{2}+\frac{1}{2}\frac{\hbar^{2}}{4JS}V_{s,1}^{2}]$ and
wavenumber $k_{s,1}=\frac{V_{s,1}}{2\beta_{0}}$ of the \textquotedblleft
carrier wave\textquotedblright\ are related by the dispersion law
$\Omega_{s,1}=\beta_{0}(k_{s,1}^{2}-A_{s,1}^{2})-2\beta_{2}=\beta_{0}%
(k_{s,1}^{2}-A_{s,1}^{2})+\frac{1}{\hbar}[g\mu_{B}B+4\left(  J^{^{\prime}%
}-J\right)  S]$. It is shown that the external magnetic field $B$ change the
frequency but not the amplitude. Then the soliton energy is seen to be
\[
E_{1}=\hbar\Omega_{s,1}=g\mu_{B}B+4\left(  J^{^{\prime}}-J\right)
S-2JSA_{s,1}^{2}+\frac{1}{2}m^{\ast}V_{s,1}^{2}%
\]
where an effective mass $m^{\ast}$ of soliton is $\frac{\hbar^{2}}{4JS}$. We
also notice that the velocity of the \textquotedblleft carrier
wave\textquotedblright, that is the phase velocity of the soliton
$\frac{\Omega_{s,1}}{k_{s,1}}=\frac{V_{s,1}}{2}-\frac{\beta_{0}A_{s,1}%
^{2}+2\beta_{2}}{k_{s,1}}$, has a negative correction $\frac{\beta_{0}%
A_{s,1}^{2}+2\beta_{2}}{k_{s,1}}$ for the half of envelope velocity, whereas
the soliton group velocity $\frac{d\Omega_{s,1}}{dk_{s,1}}=V_{s,1}$ coincides
with the envelope velocity. On the other hand when the soliton amplitude
vanishes, namely $A_{s,1}=0$, the solution $a_{1}$ in Eq. (\ref{S1}) reduces
to the linear-wave solution $a=A_{c}e^{i\varphi},$ where $\varphi
=k_{c}x-\omega_{c}t$ and group velocity $V_{c}=\frac{d\omega_{c}}{dk_{c}%
}=2\beta_{0}k_{c}$ coming from the nonlinear dispersion relation $\omega
_{c}=\beta_{0}\left(  k_{c}^{2}-2\beta_{1}^{2}A_{c}^{2}\right)  -2\beta_{2}$.

From the expression of $M_{1}$ we can directly see that $M_{1}$ is the pure
real number when $k_{c}=k_{s,1}$ and $A_{s,1}^{2}<4\beta_{1}^{2}A_{c}^{2}$.
The condition $A_{s,1}^{2}<4\beta_{1}^{2}A_{c}^{2}=\frac{4\eta(J^{^{\prime}%
}-J)}{JS}A_{c}^{2}$ is a stability criterion which is related to anisotropic
parameter $(J^{^{\prime}}-J)$ and the amplitude $A_{c}$ of the linear wave. It
is worth to point out that the condition $k_{c}=k_{s,1}$ implies the equal
group velocities for both soliton and linear wave.

\section{Two-soliton solution embedded in linear wave background}

According to the general formalism in section II it is easy to construct the
two-soliton solution in the linear wave background as follows:%

\begin{equation}
a_{2}=a_{1}+e^{i\varphi}\frac{A_{s,2}}{\beta_{1}\cosh\Gamma}e^{-i\arg h_{2}},
\label{S2}%
\end{equation}
where%
\begin{align*}
h_{2}  &  =\frac{\left(  \lambda_{2}+\overline{\lambda}_{1}\right)
\exp\left(  i\varphi\right)  -\left(  \lambda_{1}+\overline{\lambda}%
_{1}\right)  \rho_{1}+\left(  \lambda_{2}-\lambda_{1}\right)  \left\vert
\rho_{1}\right\vert ^{2}\rho_{2}}{\left(  \lambda_{2}-\lambda_{1}\right)
+\left(  \lambda_{2}+\overline{\lambda}_{1}\right)  \left\vert \rho
_{1}\right\vert ^{2}+\left(  \lambda_{1}+\overline{\lambda}_{1}\right)
\overline{\rho}_{1}\rho_{2}},\\
\Gamma &  =\ln\left\vert h_{2}\right\vert ,\text{ }\rho_{n}=\frac{L_{n}%
-\beta_{1}A_{c}e^{\theta_{n}-i\Phi_{n}}}{-\beta_{1}A_{c}+L_{n}e^{\theta
_{n}-i\Phi_{n}}},
\end{align*}
with the notations%
\begin{align*}
\theta_{n}  &  =2\operatorname{Im}M_{n}\left(  x-V_{1,n}t\right)  -x_{0,n},\\
\Phi_{n}  &  =2\operatorname{Re}M_{n}\left(  x-V_{2,n}\right)  t-\varphi
_{0,n},\\
V_{1,n}  &  =\frac{\operatorname{Im}\left(  M_{n}\gamma_{n}\right)
}{\operatorname{Im}M_{n}},V_{2,n}=\frac{\operatorname{Re}\left(  M_{n}%
\gamma_{n}\right)  }{\operatorname{Re}M_{n}},\\
\varphi &  =k_{c}x-\omega_{c}t.
\end{align*}
here the parameters $\lambda_{n}=-\mu A_{s,n}/2+ik_{s,n}/2$, $L_{n}%
=\lambda_{n}-ik_{c}/2-iM_{n}$, $M_{n}=\frac{1}{2}\sqrt{(k_{c}+i2\lambda
_{n})^{2}+4\beta_{1}^{2}A_{c}^{2}}$ and $\gamma_{n}=\left(  k_{c}%
-i2\lambda_{n}\right)  \beta_{0}$. The parameters $x_{0,n}$ and $\varphi
_{0,n}$ represent the initial center position and initial phase, which are
determined by $x_{0,n}=-\ln\left\vert C_{2,n}/C_{1,n}\right\vert $, and
$\varphi_{0,n}=\arg\left(  C_{2,n}/C_{1,n}\right)  $, respectively, where
$C_{1,n},C_{2,n}$ are the arbitrary complex constants with $n=1,2$. With the
similar reason as in the case of one-soliton solution, we often set
$C_{1,n}=1$ and $C_{2,n}$ are the arbitrary complex constants.

As the linear wave amplitude vanishes $A_{c}=0,$ we have the general
two-bright soliton solution from Eq. (\ref{S2}) in the form%
\begin{equation}
a_{2-\text{sol}}=\frac{2}{\beta_{1}\Delta_{2}}(G_{1}e^{i[k_{s,2}x-\Omega
_{s,2}t+\varphi_{0,2}]}+G_{2}e^{i[k_{s,1}x-\Omega_{s,1}t+\varphi_{0,1}]}),
\label{Br2}%
\end{equation}
where%
\begin{align*}
G_{1}  &  =\left[  \left(  \zeta_{1}-\operatorname{Re}\zeta_{3}\right)
\cosh\theta_{1}+i\operatorname{Im}\zeta_{3}\sinh\theta_{1}\right]  ,\\
G_{2}  &  =\left[  \left(  \zeta_{2}-\operatorname{Re}\zeta_{3}\right)
\cosh\theta_{2}-i\operatorname{Im}\zeta_{3}\sinh\theta_{2}\right]  ,\\
\Delta_{2}  &  =\zeta_{4}\cosh\theta_{1}\cosh\theta_{2}\\
&  -A_{s,1}A_{s,2}\left[  \cosh\left(  \theta_{1}+\theta_{2}\right)
+\cos\left(  \Phi_{1}-\Phi_{2}\right)  \right]  ,
\end{align*}
here $\zeta_{1}=A_{s,2}\left\vert \lambda_{2}+\overline{\lambda}%
_{1}\right\vert ^{2}$, $\zeta_{2}=A_{s,1}\left\vert \lambda_{2}+\overline
{\lambda}_{1}\right\vert ^{2}$, $\zeta_{3}=A_{s,1}A_{s,2}\left(
\overline{\lambda}_{2}+\lambda_{1}\right)  $, $\zeta_{4}=2\left\vert
\lambda_{2}+\overline{\lambda}_{1}\right\vert ^{2}$. The solution
$a_{2-\text{sol}}$ in Eq. (\ref{Br2}) is the general form of two-bright
soliton solution for Eq. (\ref{spin1}) which describes the interaction of two
one-bright soliton solutions with the maximal amplitudes $\frac{A_{s,n}}%
{\beta_{1}}$, the width $\frac{1}{A_{s,n}}$, envelope velocity $V_{s,n}%
=2\beta_{0}k_{s,n}$, $n=1,2$, respectively. The frequency $\Omega_{s,n}%
=\frac{1}{\hbar}[g\mu_{B}B+4\left(  J^{^{\prime}}-J\right)  S-2JSA_{s,n}%
^{2}+\frac{1}{2}\frac{\hbar^{2}}{4JS}V_{s,n}^{2}]$ and wavenumber
$k_{s,n}=\frac{V_{s,n}}{2\beta_{0}}$ of each \textquotedblleft carrier
wave\textquotedblright\ are related by the dispersion law $\Omega_{s,n}%
=\beta_{0}\left(  k_{s,n}^{2}-A_{s,n}^{2}\right)  -2\beta_{2}$. Then the
energy of each soliton is seen to be
\[
E_{n}=\hbar\Omega_{s,n}=g\mu_{B}B+4\left(  J^{^{\prime}}-J\right)
S-2JSA_{s,n}^{2}+\frac{1}{2}m^{\ast}V_{s,n}^{2},
\]
with $n=1,2$, where $m^{\ast}=\frac{\hbar^{2}}{4JS}$ denotes the effective
mass of soliton. We also notice that the velocity of each \textquotedblleft
carrier wave\textquotedblright, that is the phase velocity of each soliton
$\frac{\Omega_{s,n}}{k_{s,n}}=\frac{V_{s,n}}{2}-\frac{\beta_{0}A_{s,n}%
^{2}+2\beta_{2}}{k_{s,n}}$, has a negative correction $\frac{\beta_{0}%
A_{s,n}^{2}+2\beta_{2}}{k_{s,n}}$ for the half of envelope velocity, whereas
the group velocity of each soliton $\frac{d\Omega_{s,n}}{dk_{s,n}}=V_{s,n}$
coincides with the envelope velocity of each soliton. When the amplitude
$A_{s,n}=0,$ $n=1,2,$ the solution $a_{2}$ in Eq. (\ref{S2}) reduces to the
linear wave solution. Therefore, in general, the solution $a_{2}$ in Eq.
(\ref{S2}) represents the interaction of two one-soliton solution in a linear
wave background.

As the discussion for Eq. (\ref{S1}), we consider the case of $k_{c}%
=k_{s,1}=k_{s,2}$. Hence we have $M_{n}=\frac{1}{2}\sqrt{4\beta_{1}^{2}%
A_{c}^{2}-A_{s,n}^{2}}$, $n=1,2$. From this expression, we are easy to see
that the condition $A_{s,n}^{2}<4\beta_{1}^{2}A_{c}^{2}=\frac{4\eta\left(
J^{^{\prime}}-J\right)  }{JS}A_{c}^{2}$, $n=1,2$, becomes a stability
criterion which is related to anisotropic parameter $\left(  J^{^{\prime}%
}-J\right)  $ and the amplitude $A_{c}$ of the linear wave. It is also worth
to see that the condition $k_{c}=k_{s,1}=k_{s,2}$ implies the equal group
velocities both for each soliton and linear wave.

\section{Conclusion}

In this paper we obtain exact N-soliton solution for anisotropic spin chain
driven by a external magnetic field in linear wave background in terms of a
simple, straightforward Darboux transformation. As a special case the explicit
one- and two-soliton solution dressed by the linear wave corresponding to
magnon in quantum theory is obtained analytically and its property is
discussed in detail. The frequency $\Omega_{s,n}$, wavenumber $k_{s,n}$, and
the dispersion law of each \textquotedblleft carrier wave\textquotedblright%
\ are also studied. We obtain explicitly the energy $E_{n}$ of each soliton
and the effective mass $m^{\ast}$ of soliton. Our result show that the
stability criterion of soliton is related with anisotropic parameter and the
amplitude of the linear wave.

\section{Acknowledgment}

This work was support by the NSF of China under Grant No. 10075032.

\end{document}